\documentclass[prc,a4paper,preprint,byrevtex,onecolumn]
{revtex4}
\everymath={\displaystyle}
\usepackage{rotating}
\def\1{\mbox{l\hspace{-0.53em}1}}

\def\sig{{\bf \sigma}}
\def\sig_1{{\bf \sigma_1}}

\def\beq{\begin{equation}}
\def\eeq{\end{equation}}
\def\bea{\begin{eqnarray}}
\def\eea{\end{eqnarray}}

\begin{document}
\title{Skyrme density functional description of the double magic $^{78}$Ni nucleus}

\author{D. M. \surname{Brink $^{a}$}}
\email[E-mail: ]{thph0032@herald.ox.ac.uk}
\author{Fl. \surname{Stancu} $^{b}$}
\email[E-mail: ]{fstancu@ulg.ac.be}
\affiliation{$^{a}$ University of Oxford, Rudolf Peierls Centre of Theoretical Physics , 
1 Keble Road, Oxford OX1 3NP, U.K. \\
$^{b}$ University of Li\`ege, Institute of Physics, B.5, Sart Tilman,
B-4000 Li\`ege 1, Belgium.}

\date{\today}

\begin{abstract}
We calculate the single particle spectrum of the double magic nucleus $^{78}$Ni 
in a Hartree-Fock approach using the Skyrme density-dependent effective interaction  
containing central, spin-orbit and tensor parts. 
We show that  the tensor part has an important effect
on the spin-orbit splitting of the proton $1f$ orbit which may explain the survival of magicity
so far from the stability valley. We confirm the inversion of the $1f5/2$ and $2p3/2$ 
levels at the neutron number 48 in the Ni isotopic chain expected from
previous Monte Carlo shell model calculations and supported by experimental observation.

\end{abstract}

\keywords{Evolution of nuclear shells, tensor force, Hartree-Fock with Skyrme
interaction}

\maketitle

\section{Introduction}
Since more than a decade
there is a growing interest into the neutron- or proton-rich nuclei far from the stability
valley and into the evolution of nuclear shells in these regions. In particular
$^{78}${Ni} was expected to be one of the most neutron-rich doubly magic nucleus.
Its half-life time of 122.2(5.1)ms \cite{Xu:2014hla} and the prediction of a first excited 
state above 2 MeV \cite{Nowacki:2016isq}  were a hint of
stability of the Z = 28 and N = 50 shells. The recent experiments 
of in-beam $\gamma$-ray spectroscopy 
at the Radioactive Isotope Beam Factory of RIKEN producing the nucleus $^{79}$Cu \cite{Olivier:2017oqr},
indicated that the gaps at  Z = 28 and N = 50 remain large, which is a clear sign of stability.
At the same time the production of copper isotopes $^{75-79}$Cu at the CERN-ISOLDE 
facility supports the doubly magic character of $^{78}${Ni} \cite{Welker:2017eja}.
The magnetic dipole and the electric quadrupole moments of $^{78}_{29}$Cu and other Cu isotopes, 
measured using the CRIS experiment at the CERN-ISOLDE facility suggests that the magicity 
of Z = 28 and N = 50 is restored towards $^{78}${Ni} \cite{deGroote:2017vkw}.

The shell structure and the existence of magic numbers are a consequence of the spin-orbit interaction
\cite{Mayer:1949pd,Haxel:1949fjd}.  
Since 2005 there is much concern about the role of the tensor force
in the shell evolution and the structure of exotic nuclei, both in the framework 
of the shell model  \cite{Otsuka:2005ra,Otsuka:2005zz,Brown:2006cc,Otsuka:2006zz} 
and the Hartree-Fock Skyrme energy density functionals \cite{Colo:2007cwc}.

In a mean field approach, which leads to a one-body
potential containing a central part and a spin-orbit part,  the origin of the 
spin-orbit interaction can be clearly understood. 
In spin-saturated
nuclei the spin-orbit part stems from the spin-orbit 
nucleon-nucleon interaction.
In spin unsaturated nuclei there are additional contributions coming 
both from the exchange part of the central two-body force and from
the tensor force \cite{Stancu:1977va,Vautherin:1971aw,Beiner:1974gc}.

In an early work \cite{Stancu:1977va} we  estimated the contribution of the 
tensor part of the Skyrme interaction to the Hartree-Fock spin-orbit
splitting in several spin-saturated magic nuclei and adjusted the strength of the tensor
force such as to obtain a good global fit. 

In Ref. \cite{Brink:2007it} 
we extended the 
previous study to exotic nuclei, most of which were unknown
in 1977 and tried to shed a new light on the previous results.
We presented results for single particle levels of Sn isotopes,
N = 82 isotones and Ca isotopes, where the tensor force considerably 
improves the agreement with the experiment when its parameters
are properly chosen. 

About ten years ago the Ni isotopes were analyzed in Ref. \cite{Lesinski:2007ys}. 
There it was claimed that the currently used central and spin-orbit parts of the Skyrme 
energy density functional are not flexible enough to allow for the presence of large tensor terms.
However, ten years later, in Ref. \cite{Sushenok:2017xzn}, based on the energy density functional
of the Skyrme interaction with a tensor term and including the effect of unpaired nucleons
on the superfluid properties of nuclei,
the $\beta$-decay of $^{72-80}${Ni} isotopes were calculated and found that the $\beta$-decay half-lives 
of these neutron-rich nuclei were in reasonable good agreement with the experiment.

With this incentive, here we calculate the single particle spectrum of  $^{56-78}${Ni} isotopes 
in order to better understand the role of the tensor part and
the behaviour of the gap in the proton $1f$ shell. We found that there is an inversion 
of the $1f5/2$ and $2p3/2$ levels at $N$ = 48 consistent with the experimental proposal
of Ref. \cite{Olivier:2017oqr} and shell model calculations.
We have analyzed the behaviour of the 
neutron $1g9/2$  subshell in the Ni isotopic chain. 
Agreement was found with 
recent shell model calculations which predicted that the size of the shell gap at N = 50 is smaller
than that at N = 45 \cite{Sahin:2017kje}.  

In the next Section  we recall the original form of the tensor part of the Skyrme interaction.
In Section \ref{longrange} we remind the relation to a long range tensor force. 
In Section \ref{parameters} we introduce the parameters. In Section \ref{ni_isotopes} 
we present the calculated single particle spectra of Ni isotopes with $N$ = 40 - 50
and compare the results with other studies. The last section is devoted to conclusions.

\section{The tensor part of the Skyrme interaction}\label{tensorinteraction}

As in Ref. \cite{Stancu:1977va}, in the configuration space the tensor interaction has the following form
\begin{eqnarray}\label{tensorSkyrme}
V_{T} = \frac{1}{2} T \{ [(\vec{\sigma_1} \cdot \vec{k'})  
(\vec{\sigma_2} \cdot \vec{k'}) - \frac{1}{3} k'^2 (\vec{\sigma_1} \cdot
\vec{\sigma_2})] \delta{(\vec{r_1}-\vec{r_2})}  \nonumber \\
+  \delta{(\vec{r_1}-\vec{r_2})} [(\vec{\sigma_1} \cdot \vec{k}) 
(\vec{\sigma_2} \cdot \vec{k})      
 - \frac{1}{3} k^2 (\vec{\sigma_1} \cdot
\vec{\sigma_2})] \} \nonumber \\
+U \{(\vec{\sigma_1} \cdot \vec{k'}) \delta{(\vec{r_1}-\vec{r_2})}
 (\vec{\sigma_2} \cdot \vec{k})
 -\frac{1}{3} (\vec{\sigma_1} \cdot \vec{\sigma_2})
 [\vec{k'} \cdot \delta{(\vec{r_1}-\vec{r_2})} \vec{k}] \}.  
\end{eqnarray}
The parameters $T$ and $U$ measure the strength of the tensor force in 
even and odd states of relative motion.

The parameters of the Skyrme interaction without tensor force were originally determined 
in Hartree-Fock calculations to reproduce the 
total binding energies and charge radii of closed-shell nuclei \cite{Vautherin:1971aw}. 
Further extensive calculations were made 
later \cite{Beiner:1974gc}. Several improved parameter sets were found.    
They differ mainly  through the single particle spectra. In the present paper 
as in our previous work, we shall use the parameter set SIII which
gives good overall single particle spectra. In Ref. \cite{Stancu:1977va} a tensor 
force was added and a range of its strength was found such as to maintain a
good quality of the single particle spectra of $^{48}$Ca,$^{56}$Ni, $^{90}$Zr
and $^{208}$Pb.

Both the central exchange and the tensor interactions
give contributions to the binding energy and the spin-orbit single particle potential to be added
to the usual spin-orbit interaction. First we need to introduce the spin density
$J_{q}$ 
where $q = n,p$ stands for neutrons and protons respectively. One has \cite{Vautherin:1971aw}
\begin{equation}
J_{q} = \frac{1}{4 \pi r^3} \sum_{k}  n_{q,k}(2 j_{q,k} + 1) 
[j_{q,k}(j_{q,k} + 1) - \ell_{q,k}(\ell_{q,k} + 1) - \frac{3}{4}] R^2_{q,k}(r),
\label{spindensity}
\end{equation}
where ${k}$ runs over all occupied neutron or proton states,
$R_{k}(r)$ is the radial single particle wave function and 
$n_{q,k}$ is the occupation probability.
When the orbit is completely filled one has $n_{q,k}$ = 1.

In terms of $J_q$ the additional contribution of the central and tensor parts 
to the spin orbit potential is \cite{Stancu:1977va}
\begin{equation}
\Delta W_n = (\alpha J_n + \beta J_p) \vec{\ell} \cdot \vec{s}
\label{Wn}
\end{equation}
\begin{equation}
\Delta W_p = (\alpha J_p + \beta J_n) \vec{\ell} \cdot \vec{s}
\label{Wp}
\end{equation}
with 
\begin{equation}
\alpha = \alpha_T + \alpha_c, ~~~  \beta = \beta_T + \beta_c.
\label{alfabeta}
\end{equation}
For the Skyrme SIII interaction used in the present work the parameters of the central exchange part 
are \ \cite{Beiner:1974gc}
\begin{equation}\label{central}
\alpha_c = \frac{1}{8} (t_1 - t_2) = 61.25~ \mathrm{MeV~ fm^5},
~~~~~~\beta_c = 0~,
\end{equation}
where $t_1$ and $t_2$ are two of the Skyrme interaction parameters.
In terms of the tensor parameters $T$ and $U$ introduced in  Eq. (\ref{tensorSkyrme}) one has 
\begin{equation}\label{totaltensor}
\alpha_T = \frac{5}{12} U, 
~~~~~~\beta_T = \frac{5}{24} (T + U).
\end{equation}
Equations (\ref{Wn}) and (\ref{Wp}) imply that the mechanism invoked by 
Otsuka et al. \cite{Otsuka:2005ra,Otsuka:2005zz,Otsuka:2006zz} is intrinsic to the Skyrme energy density formalism. These 
equations show that the filling of proton (neutron) levels influences the spin-orbit
splitting of neutron (proton) levels whenever $\beta \neq 0$. In the Skyrme energy density approach
this mechanism is very simple. 

The usual spin-orbit single particle potential resulting from the two-body spin-orbit is
\begin{equation}
V_{so} =  W_0 \frac{1}{r} (\frac{d \rho}{d r} + 
\frac{d \rho_q }{d r}) \vec{\ell} \cdot \vec{s}\qquad {\rm with }\qquad \frac{d \rho}{dr}<0.
\end{equation}
The additional contributions from Eqs. (\ref{Wn}) and (\ref{Wp}) imply that
when $\beta$ is positive the neutron (proton) spin-orbit splitting is reduced as 
protons (neutrons) fill a $j = l+1/2$ level because $J_{p(n)} >0$.

It is worth mentioning that with the Skyrme density formalism, one can easily study
the combined contribution of the central exchange (\ref{central}) and tensor (\ref{totaltensor})
nucleon-nucleon interactions to the spin-orbit potential.

\section{The relation to a long range tensor force}\label{longrange}

Otsuka et al. \cite{Otsuka:2005ra} have pointed out that the nucleon-nucleon tensor force has a 
rather long range, reason for which the use of an energy density part due to the tensor force
in the Skyrme approach may not be justified. 

Equation (\ref{tensorSkyrme}) shows that the tensor term of the Skyrme interaction contains a $\delta $-function 
in the internucleon separation multiplied by momentum dependent terms. But
the momentum dependence takes the finite range of the interaction into account. 
Contrary to the view that it plays a minor role because of its $\delta$-type 
structure  \cite{Otsuka:2005ra}, this interaction
has the same effect as a finite size interaction, due to its 
momentum dependence.

In Ref. \cite{Brink:2007it}
we have shown that the expressions
(\ref{Wn}) and  (\ref{Wp}), can be used to study 
the contribution of 
finite range tensor forces.
We have used a factorization of the spin-density matrix for spherical nuclei
introduced by Negele and Vautherin \cite{Negele:1972zp}
which lead to a simplified form for a short range
tensor interaction. On the other hand we considered a tensor interaction with
a range of the order of the one pion exchange potential and calculated the 
ratio of the two contributions, say $S^Y$. 
In this way we have shown that the 
exact matrix elements of the one-pion exchange tensor potential for orbits 
with the largest $\ell$
could be expressed as a product of the short range expression given by Eq. (7) 
of Ref. \cite{Brink:2007it} and
a suppression factor $S^Y \approx 0.147$ which is almost constant for nuclei 
with mass number $A\geq 48$. It is only slightly larger, 
$\emph{i. e.}~ S^Y \approx 0.16$ for nuclei near $^{28}$Si.
Thus the short range formulae 
(\ref{Wn}) and (\ref{Wp}) with constant $\alpha$ and $\beta$  should 
give qualitatively good results   for a Yukawa one-pion exchange potential. 
One should clearly make a difference between a zero-range tensor interaction
and the tensor Skyrme interaction which is in fact finite range, as subsequently
stressed in Ref. \cite{Sagawa:2014cxa}.

Interestingly, in Ref. \cite{Lopez-Quelle:2018pan} a reduction of the strength
of the pion exchange tensor force from experimental nucleon-nucleon scattering
was found necessary to get closer to experiment for Ca and Sn isotopic chains
in a relativistic Hartree-Fock + Bardeen Cooper Schriffer (HF + BCS) approach.

Shell gaps are mainly determined by the spin-orbit splitting of the states with 
highest $l$ in any shell and our study was restricted to these states. The spin-orbit 
splitting is less important in states with lower $l$ because it is hidden by 
pairing effects and other forms of configuration mixing.

The conclusion was that the Skyrme energy functional with the tensor force
is adequate to describe the evolution of shell effects.

\section{Parameters}\label{parameters}

\begin{figure}
\includegraphics[width=12cm]{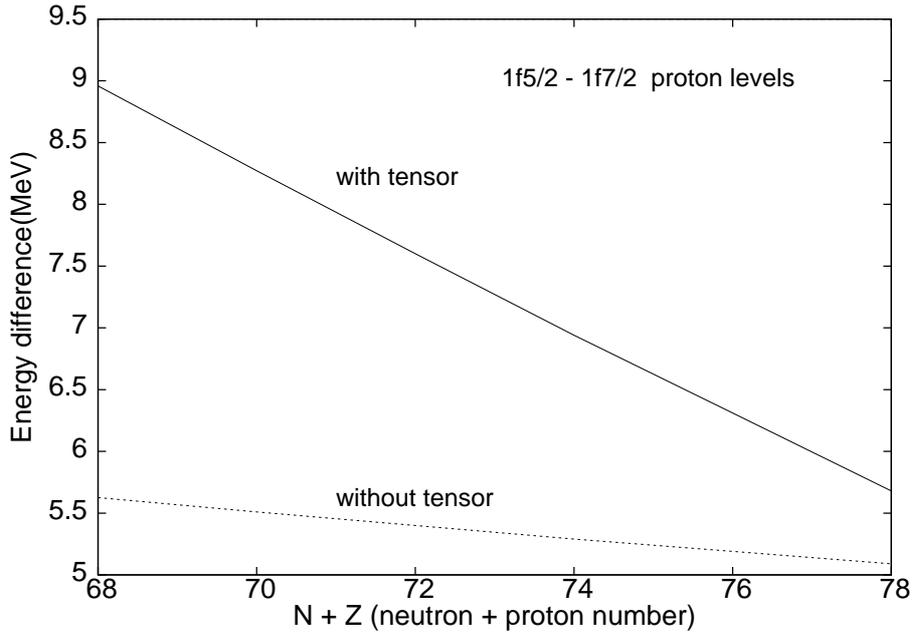} 
\caption{Proton single particle energy 
difference between the unoccupied $1f5/2$ level and the occupied $1f7/2$ level in Ni isotopes (Z = 28, N = 40 - 50)
from the Skyrme SIII interaction
without and with tensor force of parameters $\alpha$ = - 118.75 MeV fm$^5$,
$\beta$ = 120 MeV fm$^5$, Eqs. (\ref{alfabeta}).} 
\label{protongap}
\end{figure}

The considerations of the previous sections show that the simple forms
(\ref{Wn}) and (\ref{Wp}) with constant $\alpha$ and $\beta$  
are a good approximation to the contribution of the tensor forces to 
the energy density. Values of $\alpha$ and $\beta$ can be taken to be 
constant for states with maximum $l$ in nuclei with $A\geq 48$ even for 
forces with a range of the one pion exchange potential.

In Ref. \ \cite{Stancu:1977va} we searched for sets of parameters 
$\alpha$ and $\beta$ which simultaneously fit absolute values of single particle 
levels in the closed shell nuclei  $^{48}$Ca, $^{56}$Ni, $^{48}$Zr and 
$^{208}$Pb. There we found that the common optimal values were located in a 
right angled triangle 
with the sides - 80 MeV fm$^5  \leq $       $ \alpha  \leq $ 0, ~ 0 $ \leq \beta \leq $  80 MeV fm$^5$
and hypotenuse $\alpha + \beta = 0$. 
In Ref. \cite{Brink:2007it}
these constraints were relaxed because we tried to analyze single particle energies
of some  nuclei far from the stability line. 
Our choice was guided by the recent results
of Ref. \cite{Colo:2007cwc} on the Z = 50 isotopes and N = 82 isotones which
were analyzed in a HF + BCS approach based on the Skyrme interaction 
SLy5 \ \cite{Chabanat:1997un} with refitted values of $T$ and $U$
plus a pairing force.

In the present paper we still use the SIII version of the Skyrme 
interaction \ \cite{Beiner:1974gc} for comparison with the previous work.  
We maintain the conditions 
$\alpha < 0$ and $\beta > 0$      
which are not inconsistent with the previous findings 
\ \cite{Stancu:1977va}.  In Ref. \cite{Brink:2007it} we found that 
that  the values $\alpha_T$ = - 180 MeV fm$^5$
and $\beta_T$ = 120 MeV fm$^5$,  or equivalently 
$\alpha$ = - 118.75 MeV fm$^5$ and $\beta$ = 120 MeV fm$^5$, 
gave a reasonably good fit to Z = 50 isotopes and N = 82 isotones.
These values are similar to the ones fitted by Brown et al. \cite{Brown:2006cc}.


\section{Ni isotopes}\label{ni_isotopes} 

The shell gaps of the proton and neutron single particle spectra obtained in the 
present  Hartree Fock calculations with the Skyrme energy density functional
can give an indication of the double magic character of $^{78}$Ni as observed in 
the recent experimental investigation of the stability of $Z$ = 28, $N$ = 50 shells 
\cite{Olivier:2017oqr,Welker:2017eja}. Also one can study the compatibility with
large scale shell model calculations.   
An important issue is to find out to what extent the tensor part of the 
Skyrme interaction influences the stability in the case of $^{78}$Ni.
For example, Fig. \ref{protongap} shows the evolution of  the proton gap 
$e(1f5/2)$ - $e(1f7/2)$ in Ni isotopes (Z = 28, N = 40-50) 
with and without tensor force.
One can see that the effect of the tensor force is indeed important.

In both cases there is a decrease of the gap with the increase of the neutron
number. At $N$ = 40 the gap is maximum because $J_n$ = 0,
so that only the first term  in Eq. (\ref{Wp}) contributes to the spin-orbit part.
The gap 
is positive because  $\alpha J_p $ is negative  ($\alpha < 0$ and  $J_p > 0$)
as seen from the definition (\ref{spindensity}). 
For $N > 40$ both terms in Eq. (\ref{Wp}) contribute. 
As they have opposite signs due to 
$\beta > 0$, the second term reduces the contribution from the first 
and makes the gap smaller with increasing $N$, {\it i. e.} with $n_{q,k}$ 
in Eq. (\ref{spindensity}).

The decrease in the proton gap is compatible with the large scale shell model
calculation results, mentioned in Ref. \cite{Welker:2017eja},
where from the effective single particle energies it is found that the proton gap
is reduced from 6.7 MeV at $N$ = 40 to 4.9 MeV at $N$ = 50 i. e. by 1.8 MeV, due to the 
strong  $1f5/2$ - $1g9/2$ proton-neutron attractive interaction, contained in the
spin-orbit and the tensor parts. Note the recent experimental results shown in Fig. 3 of
Ref. \cite{Shand:2017mck}
attest for the first time that the proton-neutron correlations are strong enough
for a rapid change from the semi-magic structure at  $N$ = 50 to a collective structure 
at $N$ = 52. The explanation is that $Z$ = 28 is a weak sub-magic structure, as a consequence of the 
repulsive nature of the tensor force between the proton $1f7/2$ and the fully occupied neutron 
$1g9/2$. 

In our case
the reduction is of 3.28 MeV with tensor and 0.54 MeV without tensor. Thus the 
result with the tensor part included in the Skyrme interaction is closer to
the large scale shell model results. It is useful to note that
large-scale shell model calculations  including the full $p f$ shells for the protons and the full
$s d g$ shells for neutrons preserve the doubly magic nature of the ground state of $^{78}$Ni
but exhibits a well deformed prolate band at low excitation energy \cite{Nowacki:2016isq}.
Therefore, there is hope that the single particle properties are not perturbed by
complicated correlations 
which appear to be important across $Z$ = 28 and $N$ = 50 as seen from Fig. 3 of Ref.  \cite{Welker:2017eja}
describing the two-neutron separation energies.  Accordingly  the evolution of
the two-neutron shell gap as a function of the proton number seems to be an important 
observable for the strength of a shell as seen from Fig. 5  of the same paper. There is a peak at each neutron 
magic number. The overall behavior was explained in Ref. \cite{Bender:2008gi}
using a mean field calculation where the peaked structure is found to be due to quadrupole correlations.

As mentioned in Section  \ref{tensorinteraction}  the additional contribution 
brought by the tensor interaction to the spin-orbit is given by Eq. (\ref{Wp}). 
There the product $\beta J_n$ is positive because 
the parameter $\beta$ is positive in these calculations and  $J_n$ is positive
because the neutron $1g9/2$ is filled so that the proton spin-orbit splitting 
is reduced at $N$ = 50, because $\alpha J_p$ is negative, thus  weakening  the $N$ = 50 magic number.
Such a weakening has been noticed in Ref. \cite{Olivier:2017oqr} in relation to the experimental analysis of 
the $^{79}$Cu spectroscopy.

\subsection{Proton single particle spectrum}\label{SPEproton}

\begin{figure}
\includegraphics[width=12cm]{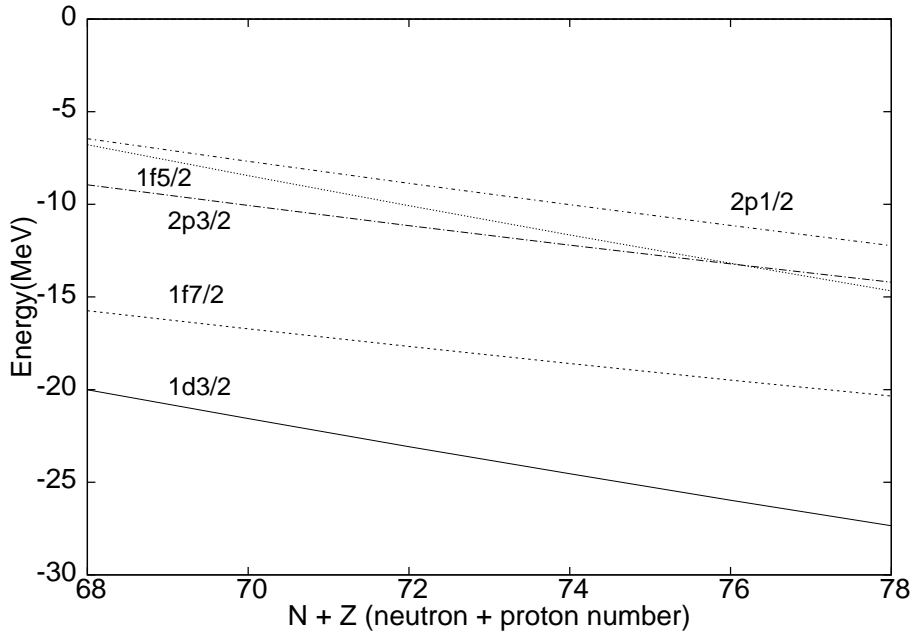} 
\caption{Proton single particle energies 
of Ni isotopes (Z = 28, N = 40 - 50) around the Fermi sea 
obtained with the Skyrme SIII interaction with the tensor force parameters $\alpha$ = - 118.75 MeV fm$^5$,
$\beta$ = 120 MeV fm$^5$.} 
\label{Niproton}
\end{figure}

\begin{figure}
\includegraphics[width=12cm]{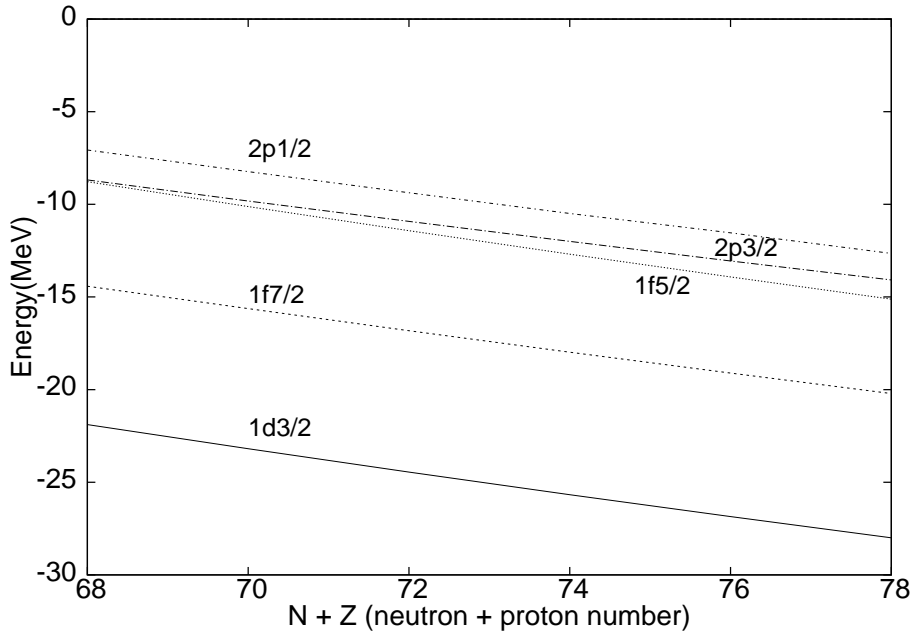} 
\caption{Proton single particle energies 
of Ni isotopes (Z = 28, N = 40 - 50) around the Fermi sea 
obtained from the Skyrme SIII interaction without tensor force, $\alpha_T$ = 0,
$\beta_T$ = 0, see Eqs. (\ref{totaltensor}).} 
\label{Niproton_notens}
\end{figure}

Comparing  Figs. \ref{Niproton} and \ref{Niproton_notens} one can see the effect of
the tensor force on the proton single particle levels around Fermi sea.   
The important difference is that while the levels $1f5/2$ and $2p3/2$ cross at $N$ = 48
when the tensor is included, they never cross beyond $N$ = 40 when the tensor is removed.
The crossing is compatible with Fig. 3 of Ref. \cite{Olivier:2017oqr} where experimental systematics
of the first $3/2^-$ and $5/2^-$ states of copper isotopes for $N$ = 40 to 50 are indicated.
The experiment suggests that the crossing takes place at $N$ = 46 so that the 
ground state of $^{79}$Cu should have a spin value of 5/2. Our results with the tensor interaction
support the proposal of Ref. \cite{Olivier:2017oqr}. The experimental excited state $3/2^-$,
(see Fig. 2 of Ref. \cite{Olivier:2017oqr} for the proposed level scheme)
lies at 656 keV above the ground state while in our case it lies at 470 keV when tensor is 
included. A fine tuning of the tensor parameters $\alpha_T$ and $\beta_T$ of Eq. (\ref{totaltensor})
may improve the agreement with the experiment, which is beyond the present purpose. 
The Monte Carlo shell-model calculations  in the $p f g9/2 d5/2$ model
space with an A3DA Hamiltonian  \cite{Tsunoda:2013hsa} performed in  Ref. \cite{Olivier:2017oqr} 
give an excitation energy of 294 keV for the $3/2^-$ level and 1957 keV for the $1/2^-$ level
while for the latter we obtain 2440 keV. The second excited level experimental of $^{79}$Cu
is placed at 1511 keV. Its structure seems to be more complicated.

On the other hand
our findings agree with the proton single particle energies calculated within a shell model
with an A3DA Hamiltonian including minor corrections,
which predict that the inversion of $1f5/2$ and $2p3/2$ levels in the Nickel chain does not take place  
before $N$ = 48, as seen from Fig. 4 of Ref. \cite{Sahin:2017kje}, very much similar to ours.
The interpretation is again as due to the 
tensor force. The probability of a state to have a single particle structure
is convincingly high in the calculated low lying  spectrum of $^{77}$Cu. The 
lowest $3/2^-$ appears at 184 keV, somewhat smaller than the experimental value of 293 keV.

An inversion of in the proton occupation of the $1f5/2$ and $2p3/2$ levels in the Nickel chain is also 
observed in Fig. 4 of Ref. \cite{Welker:2017eja}, in this case between $N$ = 44 and $N$ = 46. The explanation
given there is the effect of a strong  $1f5/2$ - $1g9/2$ proton-neutron
attractive interaction whose main active components are the spin-orbit and tensor.
Our Eqs. (\ref{Wn}) and (\ref{Wp}) are consistent with such an interpretation about the
role of the tensor force.

\subsection{Neutron single particle spectrum}\label{SPEneutron}

\begin{figure}
\includegraphics[width=12cm]{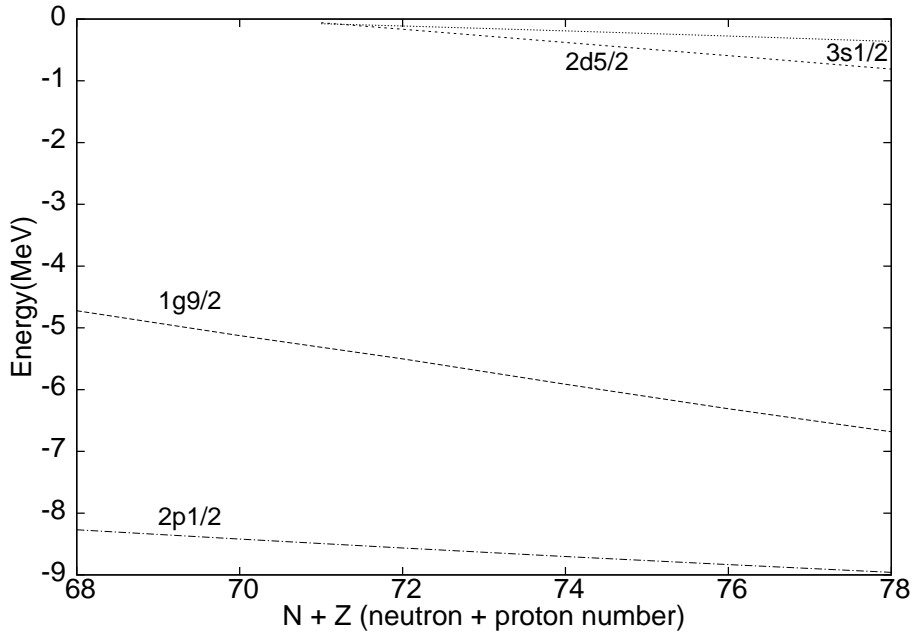}  
\caption{Neutron single particle energies 
of Ni isotopes (Z = 28, N = 40 - 50) around the Fermi sea 
from Skyrme SIII interaction with tensor force parameters $T$ and $U$ giving $\alpha$ = - 118.75 MeV fm$^5$,
$\beta$ = 120 MeV fm$^5$.} 
\label{Nineutron}
\end{figure}

\begin{figure}
\includegraphics[width=12cm]{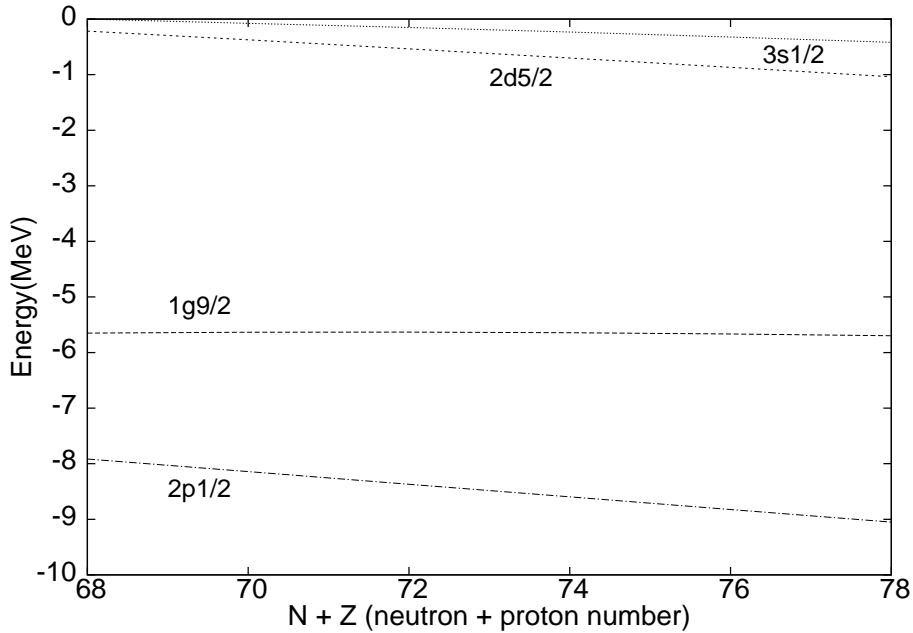}   
\caption{Neutron single particle energies 
of Ni isotopes (Z = 28, N = 40 - 50) around the Fermi sea 
calculated with the Skyrme SIII interaction without tensor force, $\alpha_T$ = 0,
$\beta_T$ = 0, see Eqs. (\ref{totaltensor}).}  
\label{Nineutron_notens}
\end{figure}

Although not much experimental information is available, 
the neutron single particle levels of Ni isotopes with $N$ = 40 - 50 around the Fermi sea have been calculated.
Fig. \ref{Nineutron} shows the result with the tensor force.
One can notice the presence of an increasingly large gap between 
the occupied $1g9/2$ and the unoccupied $2d5/2$ levels
when $N > 44$ which
takes the value of  5.87 MeV for the neutron number $N$ = 50. Note that at $N$ = 40 the level
$2d5/2$ is unbound.
Thus the stability with increasing  $N$ is larger and larger when tensor interaction is included
at variance with the hint of possible weakening of the magic number $N$ = 50 mentioned in Ref. \cite{Nowacki:2016isq}.
Such a weakening appears only when  there is no tensor contribution, see
Fig. \ref{Nineutron_notens}, where the gap decreases
from 5.43 Mev at $N$ = 40 to 4.66 MeV at $N$ = 50.
Note that when the tensor is missing the level $1g9/2$ remains practically constant from $N$ = 40
to  $N$ = 50.

\section{Conclusions}

We have performed Hartree-Fock calculations for the single particle proton and neutrons spectra for 
the Ni isotopic chain $Z$ = 28, $N$ = 40 - 50 by using the Skyrme energy density functional with
the a previously determined parametrization including a tensor term. We have found that the tensor term 
is crucial in obtaining the inversion of the $1f5/2$ and $2p3/2$ proton levels around $N$ = 48. This 
supports the doubly magic character of $^{78}$Ni  as observed in  recent experiments 
\cite{Olivier:2017oqr,Welker:2017eja} and the conclusion of Ref. \cite{Olivier:2017oqr} that
$^{79}$Cu can be described as a  $^{78}$Ni core plus a valence proton. Our calculations 
are in agreement with large scale shell model calculations which include a tensor interaction, 
as for example those of Ref. \cite{Sahin:2017kje}. The single particle spectra present a large gap both for
protons and neutrons the size of which is increased and governed by the tensor force.
The Skyrme energy density functional remains a simple, reliable and predictive approach to study the evolution 
of nuclear shells far from the stability valley.

\vspace{2cm}

\centerline{\bf Acknowledgments}

We thank Ileana Guiasu for a careful reading of the manuscript.
F.S. acknowledges support from the Fonds de la Recherche Scientifique - FNRS under 
Grant No. 4.4501.05.



\end{document}